\documentstyle[prd,aps,epsfig]{revtex}
\newcommand\hqs{\hat q^2}
\begin{document} 
           \csname @twocolumnfalse\endcsname
\title{Form factors of exclusive {\boldmath $b\to u$} transitions} 
\author{Michael Beyer$^a$ and
Dmitri Melikhov$^{b,}$\footnote{On leave of absense from Nuclear
Physics Institute, Moscow State University, Moscow, 119899, Russia}}
\address{$^a$Physics Department, Rostock University, D-18051 Rostock, Germany\\
$^b$ LPTHE,
Universit\'e de Paris XI,
B\^atiment 211, 91405 Orsay Cedex, France}
\maketitle
\begin{abstract} 
We present the form factors of the $B\to\pi,\rho$ transitions
induced by the $b\to u$ quark currents at all kinematically
accessible $q^2$. Our analysis is based on the spectral
representations of the form factors within the constituent quark
picture: we fix the soft meson wave functions and the constituent
quark masses by fitting $A_1(q^2)$ and $T_2(q^2)$ to the lattice
results at small recoils ($17\lesssim
q^2\lesssim20\;$GeV$^2$). We then calculate the $B\to\pi,\rho$
transition form factors down to $q^2=0$. For the $B\to\pi$ case the
region $q^2\lesssim 20$ GeV$^2$ however does not cover the whole
kinematically accessible range. Due to the smallness of the pion 
mass the region of small recoils is close to the nearby
$B^*(5234)$ resonance. We develop a parametrization which 
includes the $B^*$ dominance of the form factors $f_+$ and $f_-$ 
at small recoils and numerically reproduces the results of 
calculations at $q^2\lesssim 20\;$GeV$^2$.  We find
$\Gamma(B\to\pi\ell\nu)=8.0^{+0.8}_{-0.2} |V_{ub}|^2 $ps$^{-1}$ and
$\Gamma(B\to\rho\ell\nu)=15.8\pm 2.3 |V_{ub}|^2 $ps$^{-1}$.  
\end{abstract}
\vskip1cm
First measurements of the semileptonic (SL) $B\to (\pi,\rho)\;\ell\nu$
branching fractions by CLEO \cite{cleo1,cleo2} opened a possibility to
determine $|V_{ub}|$. Precise knowledge of this element of the
Cabbibo-Kobayashi-Maskawa matrix which describes the quark mixing in
the Standard Model (SM) is necessary both for understanding the
dynamics of the SM and the origin of CP violation. However, for a
proper extraction of $|V_{ub}|$ from the SL decays one needs a
reliable knowledge of the meson transition form factors which encode
the long-distance (LD) contributions to the exclusive $b\to u$
transitions.

Various nonperturbative theoretical frameworks have been applied to
the description of the meson transition form factors induced by the
$b\to u$ weak transition: among them are the constituent quark models
\cite{wsb,isgw,jaus,orsay,stech,beyer,faustov,m1,silvano,alain}, QCD
sum rules \cite{braun,rad,ball}, lattice QCD \cite{lat}, and analytical
constraints \cite{lellouch,damir}.
 
Lattice QCD simulations provide the most fundamental nonperturbative
approach and thus should lead to the most reliable results. Still,
some restrictions remain to be solved in the context of heavy-to-light
transitions. One of them is the necessity to extrapolate the
transition form factors in the heavy quark mass from the values of
order $m_c$ utilized in the lattice approach to $m_b$.  Another
problem is that lattice calulations provide the form factors only in
a region excluding large recoils.  Therefore to obtain
form factors in the whole kinematical decay region one has to rely on
some extrapolation procedures.

QCD sum rules give a complementary information on the form factors as
they allow one to determine the form factors at not very large
momentum transfers and therefore also do not cover the whole
kinematically accessible $q^2$-range \cite{braun}.  In practice,
however, various versions of QCD sum rules give rather uncertain
predictions dependent on the technical subtleties of the particular
version \cite{braun,ball}.

Various models based on the constituent quark picture have been used
for considering meson decays (see, e.g. a talk of A. Le Yaouanc for a
detailed review \cite{alain}).  An attractive feature of the
approaches based on the concept of constituent quarks is that these
approaches provide a physical picture of the process. However, a
long-standing problem of the constituent quark model (QM) applications
to meson decays is a strong dependence of the predictions on the QM
parameters.

Although none of these approaches is able at the moment to provide the
form factors in the whole accessible kinematical region of the $B$
decay, a combination of different approaches might be fruitful.  For
instance, in Ref. \cite{lat} a simple lattice-constrained parametrization
based on approximate relations obtained within the constituent quark
picture \cite{stech} and pole dominance have been proposed.  However,
within this approach the $B$ meson decays induced by the different
quark transitions, e.g. $b\to u$ and $b\to s$, remain largely
disconnected.  In Ref. \cite{mns} it was noticed that determining the soft
meson wave functions by matching the quark model calculations of the
transition form factors to the lattice results at small recoils allows
one to connect many decay processes to each other.  In this letter we
apply such an approach to a study of the $B\to\pi,\rho$ transition
form factors.

Namely, we fix the meson soft wave functions and the constituent quark
masses by fitting the lattice results to the form factors $A_1(q^2)$
and $T_2(q^2)$ at small recoils \cite{lat}, and then calculate the
form factors in the region $0< q^2\lesssim 20\;$GeV$^2$
through the spectral representations of the quark model \cite{m1}.
These spectral representations respect rigorous QCD constraints in the
limit of heavy meson decays both to heavy and light mesons and thus we
expect them to supply a reliable continuation of the lattice results
to the lower $q^2$ region.

Thus, for the $B\to\rho$ transition we calculate the form factors at
all kinematically accessible $q^2$.  For the $B\to\pi$ case this range
given above does not cover the whole kinematically accessible region. 
Also, no lattic points are
provided for $f_+(q^2)$ above $q^2>20\;$GeV$^2$. To extrapolate the
form factors $f_+$ and $f_-$ to larger $q^2$, note that the momentum 
transfers become rather close to the $B^*(5234)$ resonance. We therefore 
propose a parametrization which takes into account the $B^*$ dominance 
in the region of small recoils and reproduces the results of calculations 
at $q^2\lesssim 20\;GeV^2$.

The form factors of interest are connected with the meson transition
amplitudes induced by the vector $V_{\mu} = \bar{q_2} \gamma_{\mu}
q_1$, axial-vector $A_{\mu} = \bar{q_2} \gamma_{\mu} \gamma^5 q_1$,
and tensor $T_{\mu\nu} = \bar{q_2} \sigma_{\mu\nu} q_1$, $q_1\to q_2$
quark transition currents as follows (see notations in Ref.
\cite{mns})
\begin{eqnarray}
\label{ffs}
<P(M_2,p_2)|V_\mu(0)|P(M_1,p_1)>&=&f_+(q^2)P_{\mu}+f_-(q^2)q_{\mu},  \nonumber \\
<V(M_2,p_2,\epsilon)|V_\mu(0)|P(M_1,p_1)>&=&2g(q^2)\epsilon_{\mu\nu\alpha\beta}
\epsilon^{*\nu}\,p_1^{\alpha}\,p_2^{\beta}, \nonumber \\
<V(M_2,p_2,\epsilon)|A_\mu(0)|P(M_1,p_1)>&=&
i\epsilon^{*\alpha}\,[\,f(q^2)g_{\mu\alpha}+a_+(q^2)p_{1\alpha}P_{\mu}+
a_-(q^2)p_{1\alpha}q_{\mu}\,],   \nonumber \\
<P(M_2,p_2)|T_{\mu\nu}(0)|P(M_1,p_1)>&=&-2i\,s(q^2)\,(p_{1\mu}p_{2\nu}-p_{1\nu}p_{2\mu}), \nonumber
\\
<V(M_2,p_2,\epsilon)|T_{\mu\nu}(0)|P(M_1,p_1)>&=&i\epsilon^{*\alpha}\,
[\,g_{+}(q^2)\epsilon_{\mu\nu\alpha\beta}P^{\beta}+
g_{-}(q^2)\epsilon_{\mu\nu\alpha\beta}q^{\beta}+
g_0(q^2)p_{1\alpha}\epsilon_{\mu\nu\beta\gamma}p_1^{\beta}p_2^{\gamma}\,], 
\end{eqnarray}
where $q=p_{1}-p_{2}$, $P=p_{1}+p_{2}$. 

The dispersion approach of Refs. \cite{m1,m2} gives the transition form
factors of the meson $M_1$ to the meson $M_2$ 
as double relativistic spectral representations through
the soft wave functions of the initial and final mesons, 
$\psi_1(s_1)$ and $\psi_2(s_2)$, respectively 
\begin{eqnarray}
\label{dr}
f_i(q^2)=\int ds_1\;\psi_1(s_1)\; ds_2\;\psi_2(s_2)\tilde f_i(s_1,s_2,q^2), 
\end{eqnarray}
where $s_1$ ($s_2$) is the invariant mass of the initial (final) $\bar
qq$ pair. The double spectral densities $\tilde f_i$ of the
representation (\ref{dr}) for the $0^-\to 0^-, 1^+$ meson
decays induced by the vector, axial-vector and tensor quark currents
have been calculated in \cite{m1,m2}. The representation (\ref{dr}) is
valid for $q^2\le(m_2-m_1)^2$.

It is important to notice that the form factors (\ref{dr}) develop the
correct structure of the heavy-quark expansion in accordance with QCD
in the leading and next-to-leading $1/m_Q$ orders if the soft wave
functions $\psi_i$ are localized in the momentum space in a region of
the order of the confinement scale.  The spectral densities for all
the form factors (\ref{ffs}) have been calculated in \cite{m2}.

The spectral representations (\ref{dr}) take into account LD
contributions connected with the meson formation in the initial and
final channels. At large $q^2$ the LD effects in the $q^2$-channel
become more essential and thus one should properly replace
\begin{equation}
f_{M_1\to M_2}(q^2)\to f_{q_1\to q_2}(q^2)f_{M_1\to M_2}(q^2),  
\end{equation}
where the quark transition form factor $f_{q_1\to q_2}(q^2)$ is
introduced that accounts for the LD effects at large $q^2$ given by
the relevant hadronic resonances and continuum states.  The form
factor $f_{q_1\to q_2}(q^2)$ equals unity at $q^2$ far below the
resonance region and contains poles at $q^2=M^2_{\rm res}$. Notice that
the particular form of the quark transition form factor does not
depend on the initial and final mesons involved but rather depends on
the set of the relevant hadronic resonances and is different for the
vector, axial-vector etc channels.

\section{$B\to\rho$ transitions}

We consider the meson wave functions and the constituent quark masses
as variational parameters\footnote{One comment on the previous
application of the dispersion quark model to meson decays is in
order. In \cite{m1} it was shown that the form factors calculated
with the QM parameters of the ISGW2 model \cite{isgw} (which differs
considerably from the ISGW2 model for the transition form factors)
provide a good description of all experimental data on semileptonic
$B$ and $D$ decays.  However, the form factors of \cite{m1} have a
much flatter $q^2$-dependence and do not match the lattice results
at large $q^2$.}  and determine them from fitting the lattice
results to reproduce $T_2(q^2)$ and $A_1(q^2)$ at $q^2=19.6$ and $17.6\;$GeV$^2$
\cite{lat} by the double spectral representations (\ref{dr}) and
assuming $f_{b\to u}=1$ in the region $q^2\lesssim 20\;$GeV$^2$.
 
The soft wave function of a meson $M\;[q(m_q)\bar q(m_{\bar q})]$ can
be written as
\begin{equation}
\label{vertex}
\psi(s) = \frac{\pi}{\sqrt{2}} \frac{\sqrt{s^2 - (m_q^2 -
m_{\bar{q}}^2)^2}} {\sqrt{s - (m_q - m_{\bar{q}})^2}} \frac{w(k^2)}{s^{3/4}}, 
\end{equation}
where $k^2=\lambda(s, m_q^2, m_{\bar{q}}^2)/4s$ with 
$\lambda(s,m_q^2,m_{\bar q}^2)=(s-m_q^2-m_{\bar q}^2)^2-4m_q^2m_{\bar q}^2$,  
and the ground-state radial $S$-wave function $w(k^2)$ 
is normalized as $\int w^2(k^2)k^2 dk=1$.
For the functions $w(k^2)$ we assume a simple gaussian form 
\begin{equation}
w(k^2)\propto\exp(-k^2/2\beta^2)
\end{equation}
where $\beta$ to be obtained by a fit. 

The ranges of the $B$ and $\rho$ are shown in Table \ref{table:parameters}. 
The values of the constituent quark masses and the slope parameter 
$\beta_\rho$ are fixed rather tightly by the $\chi^2$ fit to the lattice 
data, whereas 
$\beta_B$ cannot be fixed with a good accuracy. We determine the ranges of
$\beta_B$ such that the leptonic decay constant $f_B$ calculated through 
the relation \cite{m1} 
\begin{equation}
\label{fp}
f_P=\sqrt{N_c}(m_q+m_{\bar q})\int
ds\;\psi(s)\frac{\lambda^{1/2}(s,m_q^2,m_{\bar q}^2)} 
{8\pi^2 s}\frac{s-(m_q-m_{\bar q})^2}{s}. 
\end{equation}
lies in the interval $f_B=170\pm30 \;MeV$ in accordance with the lattice 
estimates \cite{lat}. 
Once the wave functions and the quark masses are determined, we use
the spectral representations (\ref{dr}) for calculating all the form
factors for the $B\to\rho$ transition in the whole kinematically
accessible region.  Fig. 1 illustrates the calculated form factors 
versus the lattice data. Table \ref{table:ffs} gives parameters of a 
convenient interpolation of the results of the calculation in the form
\begin{equation}
\label{inter}
f(q^2)=\frac{f(0)}
{1-\sigma_1 \hqs+\sigma_2\hat q^4},  
\end{equation}
where we have introduced $\hat q^2=q^2/M^2_{B^*}$ with 
$M_B^*=5.324\;$GeV.  Since we have calculated the form factors at all
kinematically accessible $q^2$ the particular form of the fit function
is not important.  The interpolations (\ref{inter}) deviate from the
results of calculation by less than 1\%. The calculated decay
rates are given in Table \ref{table:rates}.

\section{$B\to \pi$ transitions}

For the transition $B\to\pi$ a new wave function parameter $\beta_\pi$
appears. It is not independent and strongly correlates
with $m_u$ through $f_\pi$ given by eq (\ref{fp}).
Requiring $f_P=132\;$MeV this implicitly determines $\beta_\pi$ once
$m_u$ is fixed. 

With the wave functions gived, we calculate the $B\to\pi$ transition
form factors at $0<q^2\lesssim 20\;$GeV$^2$. The form factors
versus the lattice results shown in Fig. 1 are found to be in
perfect agreement. This confirms our assumption $f_{b\to u}=1$ at
$q^2\lesssim 20\;$GeV$^2$.  This region however does not cover the whole
kinematically accessible range. To find the form factors at larger
$q^2$ we must use some extrapolation procedure.

In the region of small recoils the form factors 
are dominated by the neighbouring $B^*$ poles and one finds  
\begin{eqnarray}
f_+(q^2)&=&\frac{g_{B^*B\pi}f_{B^*}}{2M_{B^*}(1-q^2/M^2_{B^*})}+
{\rm regular\; terms\; at}\;q^2=M^2_{B^*}, \\ 
f_-(q^2)&=&\frac{g_{B^*B\pi}f_{B^*}}{2M_{B^*}(1-q^2/M^2_{B^*})}
\frac{M^2_B-M^2_\pi}{M^2_{B^*}}+
{\rm regular\; terms\; at}\;q^2=M^2_{B^*},
\end{eqnarray}
where the $B^*B\pi$ coupling constant $g_{B^*B\pi}$ is defined through
$\langle
\pi(p_2)B^*(q)|B(p_1)\rangle=g_{B^*B\pi}\epsilon^{*}_\alpha(q)p_2^\alpha$.
The regular terms here stand for the contribution of other resonances
and continuum hadronic states. It should be noted, that both the vector $1^-$
 and scalar $0^+$ resonances contribute to $f_-$ whereas only vector $1^-$ 
states contribute to $f_+$ (see e.g. \cite{grinstein}). 

Regular terms might be taken into account by assuming a single-pole form for the form
factors with a modified $q^2$-dependent 'residue' as follows
\begin{eqnarray}
\label{fpmpar1}
f_\pm(q^2)&=&\frac{\hat f_\pm(\hat q^2)}{1-\hat q^2}, 
\end{eqnarray}
where 
\begin{eqnarray}
\hat f_+(1)&=&\frac{g_{B^*B\pi}f_{B^*}}{2M_{B^*}} \nonumber \\
\label{relation}
\hat f_-(1)&=&-\hat f_+(1)\frac{M_B^2-M_\pi^2}{M^2_{B^*}}. 
\end{eqnarray}

Using the PCAC prescription for the pion field, the 
$B^*B\pi$ coupling constant can be estimated at the unphysical point 
$g_{B^*B\pi}(p_1^2=M_B^2,q^2=M_B^2,p_2^2=0)$. At this point the coupling
constant is represented through the meson transition form factor 
$f_{P(M_B)\to V(M_B)}$ which can be calculated within the same dispersion approach. 
Namely, we find
\begin{eqnarray}
\langle \pi(p_2)V(q)|P(p_1)\rangle&=&\lim_{p_2^2\to 0, q^2\to p_1^2}
\frac{1}{f_\pi}\epsilon^*_\alpha(q)p_2^\alpha
\left[f(p_2^2,p_1^2,q^2) \right.\nonumber\\
&&\left.+a_{+}(p_2^2,p_1^2,q^2)(p_1^2-q^2)
+a_{-}(p_2^2,p_1^2,q^2)p_2^2\right]  \nonumber \\ 
&=&\frac{1}{f_\pi}\epsilon^{*}_\alpha(q)p_2^\alpha f(0,M_P^2,M_P^2), 
\end{eqnarray} 
and the form factor $f(0,M_B^2,M_B^2)$ of the $B\to B^*$ transition
is calculated through the spectral representation (\ref{dr}) assuming
identical radial wave functions of $B^*$ and $B$ mesons. In the heavy
quark limit this is a rigorous property, and we expect this
approximation to work well for real $B$ and $B^*$ mesons.
To get to the physical point $g_{B^*B\pi}(m_\pi^2,M_B^2,M^2_{B^*})$ one needs to perform 
a continuation which is not unique. However due to the small difference of the 
$B$ and $B^*$ meson masses we expect 
$g_{B^*B\pi}(m_\pi^2,M_B^2,M^2_{B^*})\simeq g_{B^*B\pi}(0,M_B^2,M^2_{B})$. 

The result of the calculation of $f(0,M_B^2,M_B^2)$ is weakly sensitive 
to the values of the quark masses but mostly depends on the $B$ wave 
function. The value $f(0,M_B^2,M_B^2)$ strongly correlates with $f_B$
such that the relation
\begin{eqnarray}
g_{B^*B\pi}=\frac{9\pm0.4\;\mbox{GeV}}{f_B}
\end{eqnarray} 
is fulfilled for the range of the QM parameters which reproduce
$f_B=170\pm30\;$MeV. 
The Sum Rule analysis of the $g_{B^*B\pi}$ and references to other results 
can be found in \cite{sr}.  

Finally, the residue of the form factor $f_+$ at the $B^*$ pole
takes the value
\begin{equation}
\label{hatfplus1}
\hat f_+(1)=(0.8\pm0.04)f_{B^*}/f_B, 
\end{equation}
and for further numerical estimates we use $f_{B^*}/f_B=1.2\pm0.1$. 

For the quantities $\hat f_\pm$ we assume a smooth parametrization
\begin{eqnarray}
\label{fplusminus}
\hat f_\pm(\hat q^2)=\frac{f_\pm(0)}
{(1-\sigma_1^{\pm}\hat q^2+\sigma_2^{\pm}\hat q^4)}, 
\end{eqnarray}
where the coefficients $\sigma_{1,2}$ are not independent:  
Eq. (\ref{hatfplus1}) gives 
\begin{equation}
\frac{f_+(0)}{1-\sigma_1^++\sigma_2^+}=(0.8\pm0.04)
\frac{f_{B^*}}{f_B} 
\end{equation}
and the relation (\ref{relation}) leads to
\begin{eqnarray}
\label{sigmaconstraint}
\frac{f_+(0)}{(1-\sigma_1^{+}+\sigma_2^+)}
+\frac{f_-(0)}{(1-\sigma_1^{-}+\sigma_2^-)}\frac{M_B^{*2}}{M_B^2-M_\pi^2}=0.
\end{eqnarray}
The parameters $\sigma_{1,2}$ are determined from the 
$\chi^2$-fit to the results of the calculation at $q^2\lesssim 20 \;$GeV$^2$.  
Table
\ref{table:ffs} presents the relevant numbers. At $q^2\ge 20\;$GeV$^2$
the parametrizations are used for extrapolation of the form factors
$f_{\pm}$ to all kinematically accessible $q^2$ (see Fig. 1).

For the form factor $f_0(q^2)=f_+(q^2)+q^2 f_-(q^2)/Pq$ a
combination of PCAC and current algebra yields the relation \cite{pcachq}
\begin{equation}
f_0(M_B^2)=f_B/f_\pi 
\end{equation} 
Using the value $f_B=170\pm 30\;$MeV we obtain
$$
f_0(M_B^2)=1.35\pm0.3
$$
which is found to be in a reasonable agreement with the results of our 
extrapolating formulas. 

The calculated $B\to\pi\ell\nu$ decay rate is gived in Table
\ref{table:rates}. Notice that the details of the high-$q^2$ behavior
of the form factors which depend on the extrapolation procedure do not
affect considerably the decay rate. The latter is mostly determined by
the region $q^2\lesssim 20\;$GeV$^2$ where the form factors are calculated
directly.

\begin{figure}
\begin{center}
\begin{tabular}{cc}
\mbox{\epsfig{file=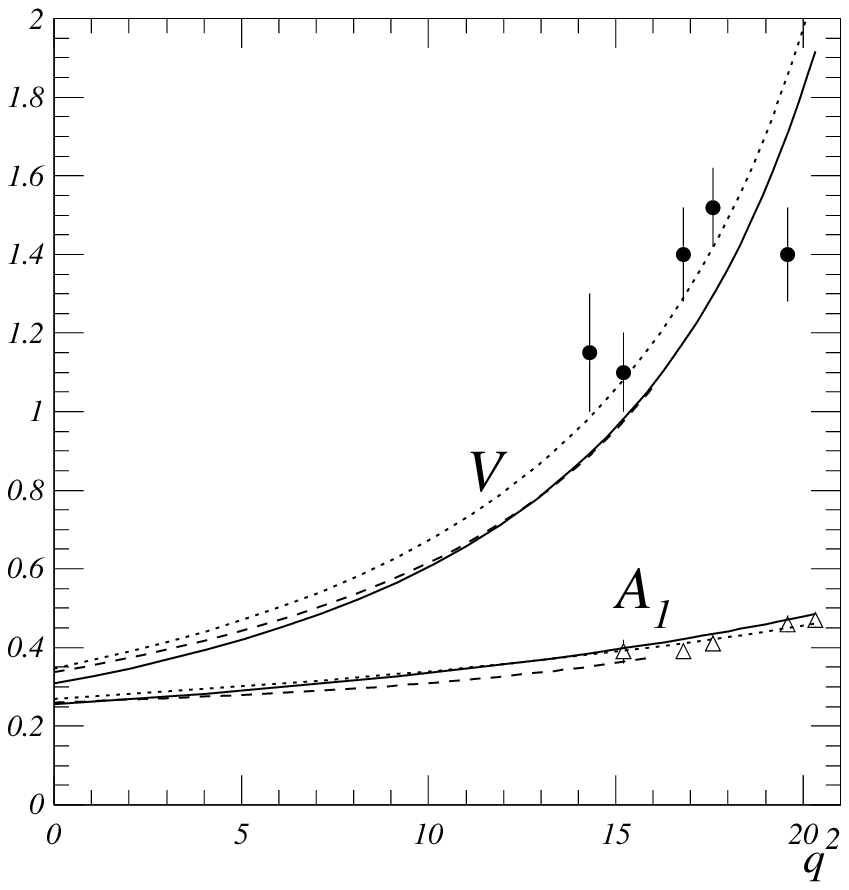,width=6.cm}}
&\mbox{\epsfig{file=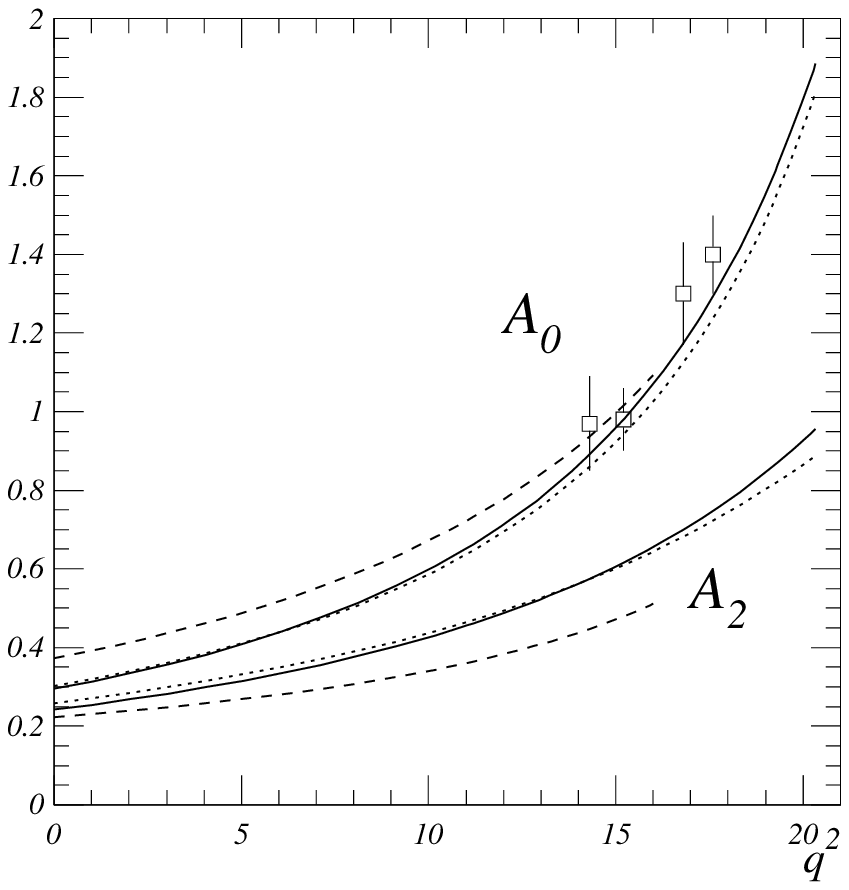,width=6.cm}}\\
\mbox{\epsfig{file=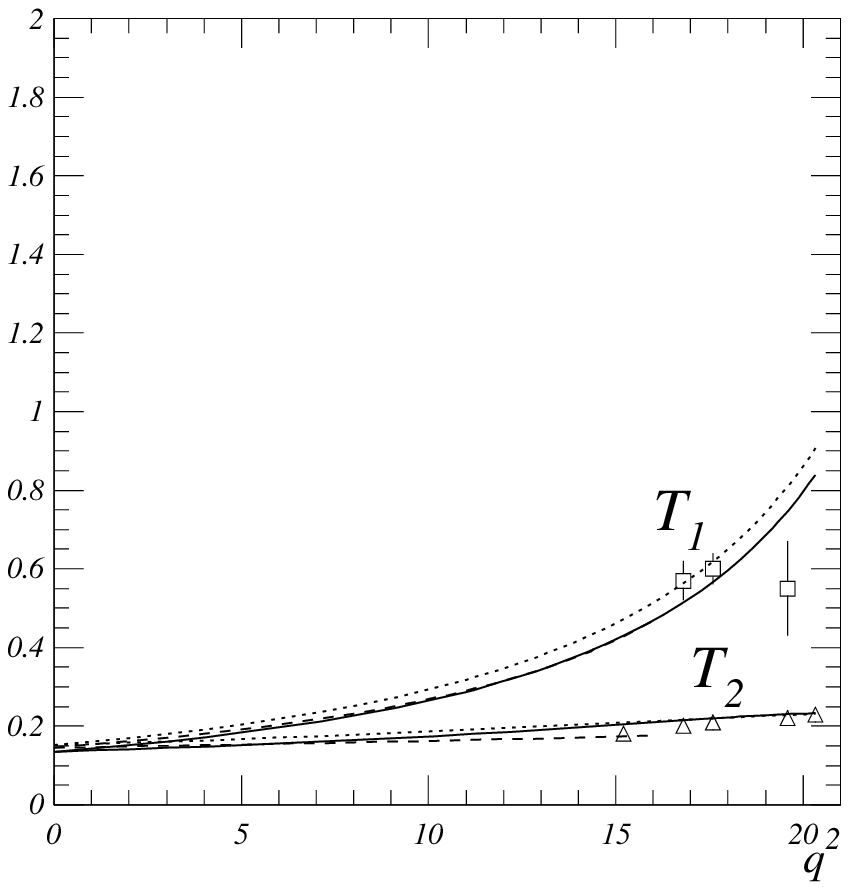,width=6.cm}}
&\mbox{\epsfig{file=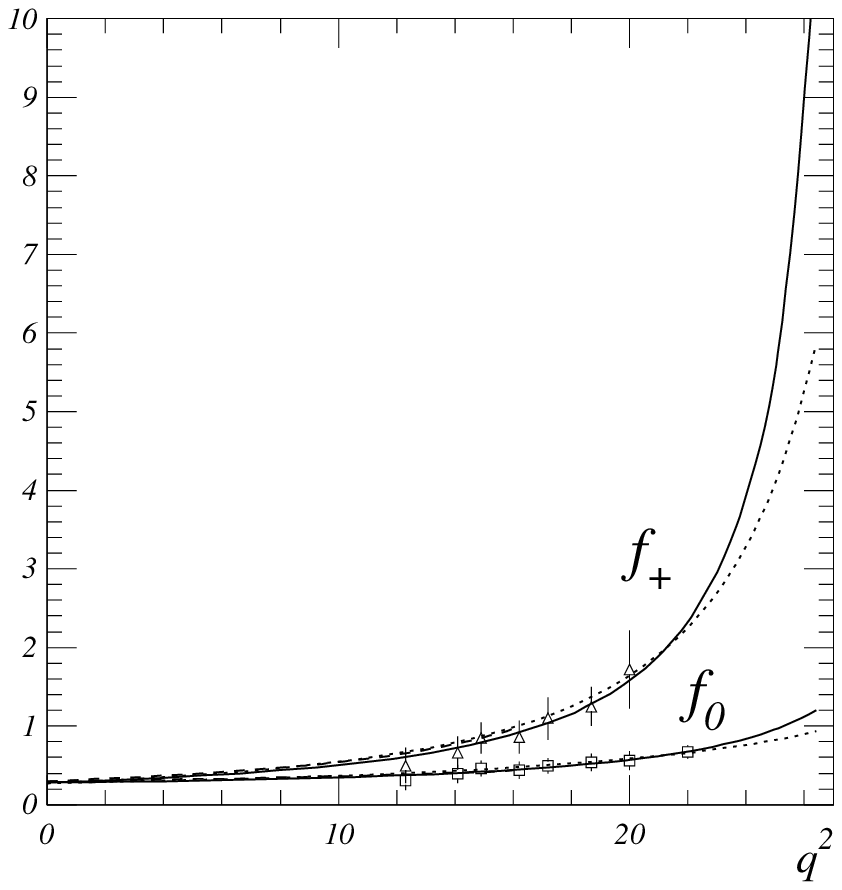,width=6.cm}}\\
\end{tabular}
\caption{\label{fig:ffs}
The form factors of the $B\to\rho$ and $B\to\pi$ transitions
through the $b\to u$ quark currents vs. lattice data \protect\cite{lat} 
and calculations within different approaches. 
$A_1=f/ (M_B + M_\rho)$, 
$A_2=-(M_B + M_\rho)a_+$, 
$A_0=[q^2 a_{-}+f+(M^2_B-M^2_\rho) a_{+}]/2M_\rho$, 
$V=(M_B + M_\rho) g$,
$T_1(q^2)=-g_+/2$, 
$T_2=-\frac{1}{2}(g_{+}+q^2 g_{-}/(M_B^2-M^2_\rho))$. 
Solid lines - our QM results, dotted lines - lattice-constrained parametrizations of
\protect\cite{lat}, dashed lines - LCSR \protect\cite{ball}.}
\end{center}
\end{figure}

Fig. \ref{fig:ffs} compares our results with recent light-cone sum
rule calculations available at $q^2\lesssim 16\;$GeV$^2$ \cite{ball} and 
lattice-constrained parametrizations of ref. \cite{lat}. One can see that the 
results of different approaches to the form factors do not differ 
significantly. However, it should be taken into account that in the case of the
$B\to\rho$ transition this minor difference in the form factors provides rather
sizeable spread of predictions for the decay rates.  

Summing up, we have analyzed the form factors of the exclusive $b\to
u$ transition using the spectral representations based on constituent
quark picture and obtain form factors in the whole kinematically
accessible region.

The meson wave functions and the constituent quark masses have been
determined by describing the results of lattice simulations of the
form factors $A_1(q^2)$ and $T_2(q^2)$ at small recoils. This allowed
us to calculate the form factors at $q^2\lesssim 20\;$GeV$^2$
which cover all kinematically accessible $q^2$ in the $B\to\rho$
transition.  In the $B\to\pi$ case the interval $q^2\lesssim
20\;$GeV$^2$ does not cover the kinematically accessible region and an
extrapolation to higher $q^2$ is necessary. To this end we take into
account the dominance of the form factors at small recoils by the
$B^*$ pole. The calculated $B\to\pi\ell\nu$ decay rate is found to be
only slightly sensitive to the particular details of the extrapolation
procedure.

We take the pleasure in thanking D. Becirevic and H. Schr\"oder for
helpful discussions which considerably influenced the final form of
the paper. We are grateful to A. Le Yaouanc and O. Pene for the
interest in this work and to DFG for financial support under grant 436
RUS 18/7/98. D.M. is grateful to the Physics Department of the Rostock
University for hospitality.

\begin{table}[1]
\caption{\label{table:parameters}
Quark masses and the slope parameters of the soft meson 
wave functions (in GeV).} 
\centering
\begin{tabular}{|c|c|c|c|c|}
$m_b$ & $m_u$ &$\beta_{B}$ & $\beta_{\pi}$ & $\beta_{\rho}$ \\ 
\hline
4.85$\pm$0.03  & 0.23$\pm$0.01 & 0.54$\pm$0.04 & 0.36$\pm$0.02  & 0.31$\pm$0.03  
\end{tabular}
\end{table}

\begin{table}[2]
\caption{\label{table:ffs}
Parameters of the fits to the calculated $B\to \pi,\rho$ transition 
form factors in the form (\ref{fpmpar1}), (\ref{fplusminus}) for $f_\pm$ and 
(\ref{inter})  for all other form factors. The numbers correspond to
the central values of the QM parameters given in Table 
\ref{table:parameters}.}  
\centering
\begin{tabular}{|c|c|c|c|c|c|}
 & $f_{+}$ & $f_{-}$  &$s$  &$g$ &$f$ \\ 
\hline
$f(0)$  & 0.284   & $-$0.247  & 0.05  & 0.051  & 1.55  \\
$\sigma_1$     & 0.184   &    0.16   & 1.5   & 1.60   & 0.69 \\
$\sigma_2$     &$-0.52$  & $-0.577$   & 0.5   & 0.60  & 0.041  \\
\hline
 &$a_{+}$ & $a_{-}$  & $g_{+}$  & $g_{-}$ & $g_{0}$ \\  
\hline
$f(0)$   & $-$0.04  &  0.044 & $-$0.27 &   0.25  & 0.00374  \\
$\sigma_1$      &   1.40   &  1.49  & 1.60    &   1.61  & 2.36     \\
$\sigma_2$      &   0.50   &  0.54  & 0.60    &   0.60  & 1.64   
\end{tabular}
\end{table}

\begin{table}[3]
\caption{\label{table:rates}
Decay rates in units $|V_{ub}|^2{\rm ps^{-1}}$.}
\centering
\begin{tabular}{|c|c|c|c|}
Ref. & $\Gamma(B\to\pi\ell\nu)$  &  $\Gamma(B\to\rho\ell\nu)$  &
$\Gamma_L/\Gamma_T$   \\ 
\hline
This work &       $8.0^{+0.8}_{-0.2}$ & $15.8 \pm 2.3$ & $0.88\pm0.08$  \\
ISGW2 QM \cite{isgw}  & 9.6 & 14.2 & 0.3   \\
Lat \cite{lat} & $8.5^{+3.4}_{-0.9}$ & $16.5^{+3.5}_{-2.3}$ &
$0.80^{+0.04}_{-0.03}$ \\ 
LCSR \cite{braun} & $-$    &   13.5$\pm$ 4.0 & 0.52$\pm$ 0.08  
\end{tabular}
\end{table}

\end{document}